\documentclass[journal]{IEEEtran}

\usepackage{graphicx}
\usepackage[cmex10]{amsmath}
\usepackage{cases}
\usepackage{amsthm} %proof
\usepackage{cite}
\usepackage{amssymb}
\usepackage{algorithm}
\usepackage{algorithmic}
\usepackage{amsmath}
\usepackage{subeqnarray}
\usepackage{cases}
\usepackage{enumerate}
\usepackage{booktabs}
\usepackage{stfloats}
\usepackage{epstopdf}
\usepackage{float}
\usepackage{esint}
\usepackage{subfigure}
\usepackage{color}
\usepackage{mathrsfs}
\usepackage[final]{changes}

\newtheorem{theorem}{\hskip\parindent\bf{Theorem}}

\ifCLASSINFOpdf
\else
\fi

\begin{document}

\vspace{-8mm}
%\title{\LARGE{Simultaneously Transmitting and Reflecting RIS (STAR-RIS) Assisted Multi-Antenna Covert Communication: Optimization and Analysis}}
%\title{STAR-RIS Assisted Multi-Antenna Covert Communications in NOMA System:  Analysis and Optimization}
\title{\LARGE{STAR-RIS Assisted Covert Communications in NOMA Systems}}

\vspace{-4mm}
\author{Han Xiao,~\IEEEmembership{Student Member,~IEEE,} Xiaoyan Hu$^*$,~\IEEEmembership{Member,~IEEE,} \\
%Yongxu Zhu,~\IEEEmembership{Member,~IEEE,}
Tong-Xing Zheng,~\IEEEmembership{Member,~IEEE,} and Kai-Kit~Wong,~\IEEEmembership{Fellow,~IEEE}
%Pengcheng Mu,~\IEEEmembership{Member,~IEEE,}   Wenjie Wang,~\IEEEmembership{Member,~IEEE,}  Tong-Xing Zheng,~\IEEEmembership{Member,~IEEE,} \\
%Pengcheng Mu,~\IEEEmembership{Member,~IEEE,}  Wenjie Wang,~\IEEEmembership{Member,~IEEE,}\\
%Kai-Kit~Wong,~\IEEEmembership{Fellow,~IEEE}, Kun~Yang,~\IEEEmembership{Fellow,~IEEE}

}
\vspace{-8mm}
\maketitle
\vspace{-8mm}
\begin{abstract}
Covert communications assisted by simultaneously transmitting and reflecting reconfigurable intelligent surface (STAR-RIS) in non-orthogonal multiple access (NOMA) systems have been  explored in this paper. In particular, the access point (AP) transmitter adopts NOMA to serve a downlink covert user and a public user.
%To enhance the covert performance, the uncertainty of the background noise at the warden is exploited.
The minimum detection error probability (DEP) at the warden is derived considering the uncertainty of its background noise, which is used as a covertness constraint. % to represent the covert performance.
We aim at maximizing the covert rate of the system by jointly optimizing AP's transmit power and passive beamforming of STAR-RIS, under the covertness and quality of service (QoS) constraints. An iterative algorithm is proposed to effectively solve the non-convex optimization problem. Simulation results show that the proposed scheme significantly outperforms the conventional RIS-based scheme in ensuring system covert performance. % of covert communications.
\end{abstract}
\maketitle
\vspace{-3mm}
\section{Introduction}\label{sec:S1}
\vspace{-1mm}
The technology of covert communications (CCs) as a new security paradigm has attracted significant research interest in both civilian and military applications\cite{yan19}. It can conceal the existence of communications between transceivers and provide a higher level of security for wireless communications than physical layer security. As a pioneer work, \cite{bash13} first establishes the fundamental limit of CCs over additive white Gaussian noise (AWGN) channels from the perspective of information theory. %, which demonstrates that $O(\sqrt n)$ bits information can be transmitted covertly and reliably from transmitters to receivers over $n$ channel uses.
%detection error probability (DEP) of warden will go to zero if the amount of transmitted information exceeds this square root law.
Actually, the inherent unpredictability of wireless channels and the interference from other sources are ignored in  \cite{bash13}, %the communication systems
leading to a pessimistic solution. %For example,
Later, \cite{he2017covert, wang18} demonstrate that %over $O(\sqrt{n})$ goeckel15,
more information bits can be covertly transmitted  when  eavesdroppers do not exactly know the power of background noise or channel state information (CSI). %to the receiver
%Besides, existing works also resort to other uncertainties, e.g., power-varying artificial noise \cite{Hu19}, uninformed jammer \cite{zheng21},  to enhance the performance of covert communications. %

The aforementioned works have validated the effectiveness of the CC techniques from different perspectives, however, they just investigate the simple CC scenarios with only one legitimate user served by a single-antenna AP transmitter. In order to expand to multi-user scenarios, multiple access techniques have to be adopted. Non-orthogonal multiple access (NOMA) is a promising technique that can achieve higher spectral efficiency, lower access delay, and massive connectivity than orthogonal multiple access (OMA) techniques \cite{ding2020unveiling}. Hence, NOMA holds significant potentials for wide-ranging applications in wireless communications and has been leveraged in CCs. For example, a covert NOMA communication scheme is proposed in \cite{tao20}, where the NOMA-weaker user transmits with random power to facilitate the covert transmissions between the covert user and the transmitter. In \cite{lv2021covert}, authors explores the CCs of both the downlink and uplink transmissions in NOMA system.
% \textcolor[rgb]{1.00,0.00,0.00}{where all the nodes are equipped with a single antenna}.

While NOMA offers lots of advantages for communication systems, it is incapable of tackling the challenges posed by the randomness of wireless channels, which highly restrain the performance gains facilitated by NOMA. %As a result, the performance gains facilitated by NOMA may be limited.
%To break through this limitation,
To desirably control the wireless propagation environment, the reconfigurable intelligent surface (RIS) and a more advanced  RIS called simultaneously transmitting and reflecting RIS (STAR-RIS) have emerged as promising solutions %Liu22,zhou21intelligent,
which has been leveraged in many wireless communication scenarios including covert communications \cite{X.HU_TCOM21RIS,liu21star,   xiao2023simultaneously}.  %X.HU_TCOM21RIS, lu20intelligent, \cite{lu20intelligent, xiao2023simultaneously}. %X.HU_TCOM21RIS,zhou21intelligent,Liu22
%Specifically, \cite{lu20intelligent} generally summarizes the application potentials of RIS in improving covert communications. %After that, some works on covert communication aided by RIS are proposed. For example,
%Later in \cite{zhou21intelligent}, authors demonstrate that RIS can enable perfect covertness for covert communications and analyze the performance gain under the assumption that the instantaneous CSI at the warden is available.
%However, a limitation of the existing works using RIS is that they only consider the reflection mode of RIS, which restricts the transmitters and receivers to be on the same side of RIS. However, in realistic scenarios, users may be located on either side of RIS, and thus the conventional RIS may not be flexible and effective enough for these cases.
%
%To overcome this limitation, a novel technology called simultaneously transmitting and reflecting RIS (STAR-RIS) is further emerged \cite{liu21star}. %xu21star,
Unlike conventional RIS, STAR-RIS offers a more flexible full-space smart radio environment with 360$^\circ$ coverage, which can simultaneously control the coefficients of reflected and transmitted signals. This feature of STAR-RIS has drawn great attention from both academia and industry for its potential applications in wireless communications. However, the investigation of STAR-RIS aided wireless communication systems is still in its infancy stage. As for secure communications, only a small number of state-of-the-art works have utilized STAR-RISs to enhance the system secure performance with NOMA techniques \cite{han22artificial,zhang22secrecy}.

In this paper, we investigate the STAR-RIS assisted \mbox{CC} in NOMA systems to exploit its potential in enhancing the system covert performance. Specifically, we derive closed-form expressions for the minimum DEP and optimal detection threshold at the warden considering the worst-case scenario. We establish an optimization problem to maximize the covert rate under CC and quality of service (QoS) constraints by jointly optimizing the AP's transmit power allocation and the STAR-RIS's passive beamforming. The problem is non-convex and challenging to solve directly due to strong coupling among optimization variables. Hence, we propose an alternating optimization algorithm to solve it iteratively by addressing two subproblems: deriving the optimal power allocation in closed form with given passive beamforming and designing reflection and transmission coefficients effectively by the SDR method with given power allocation.
%\textit{Notation:} Operator $\circ$ represents the Hadamard product. $\operatorname{Diag}(\mathbf{a})$ denotes a diagonal matrix with diagonal elements in vector $\mathbf{a}$, $\operatorname{diag}(\mathbf{A})$ denotes a vector whose elements are composed of the diagonal elements of matrix $\mathbf{A}$. $|\cdot|$ and $\|\cdot\|_2$ are the complex modulus and the spectral norm, respectively. $\mathbf{A}\succeq0$ means that  $\mathbf{A}$ is a positive semidefinite matrix. $\mathbf{I}_{N\times1}$ represents the vector with $N\times 1$ entries that are $1$.
\vspace{-3mm}
\section{System Model}\label{system_model}
\vspace{-1mm}
In this paper, we consider a STAR-RIS-assisted covert NOMA communication system, consisting of a single-antenna AP transmitter (Alice)
aided by a STAR-RIS with $M$ elements, a covert user (Bob), a public user (Carol) and a warden user (Willie) all equipped with a single antenna. We assume that the STAR-RIS is deployed at the users' vicinity to enhance the  communications between Alice and legal users, i.e., Bob and Carol, which
locate on opposite sides of the STAR-RIS  and can be simultaneously served  by the reflected (T) and transmitted (R) signals via STAR-RIS. %, respectively.
%\vspace{4mm}as shown in Fig. \ref{fig:System}
%\begin{figure}[ht]
%	\centering
%	\includegraphics[scale=0.30]{Scene.eps}
%\vspace{-1mm}
%	\caption{System model for STAR-RIS-assisted covert communications.}\label{fig:System}
%\end{figure}

The wireless communication channels from Alice to STAR-RIS, and from STAR-RIS to Bob, Carol, Willie are represented as
\textbf{h$_{\rm{AR}}$}=$\sqrt{\textit{l$_{\rm{AR}}$}}\textbf{g$_{\rm{AR}}$}$$\in\mathbb{C^{\textit{M$\times1$} }}$,
\textbf{h}$_{\rm{rb}}$=$\sqrt{\textit{l$_{\rm{rb}}$}}\textbf{g$_{\rm{rb}}$}$$\in\mathbb{C^{\textit{M$\times 1$}}}$,
\textbf{h}$_{\rm{rc}}$=$\sqrt{\textit{l$_{\rm{rc}}$}}\textbf{g$_{\rm{rc}}$}$$\in\mathbb{C^{\textit{M$\times 1$} }}$ and \textbf{h}$_{\rm{rw}}$=$\sqrt{\textit{l$_{\rm{rw}}$}}\textbf{g$_{\rm{rw}}$}$$\in\mathbb{C^{\textit{M$\times 1$} }}$, respectively.
Here, \textbf{g$_{\rm{AR}}$} and \textbf{g$_{\rm{rb}}$}, \textbf{g$_{\rm{rc}}$}, \textbf{g$_{\rm{rw}}$} are the small-scale Rayleigh fading coefficients with independent identically distributed (i.i.d.) entries following complex Gaussian distribution with zero mean and unit variance.
In addition, \textit{l$_{\rm{AR}}$} and \textit{l$_{\rm{rb}}$}, \textit{l$_{\rm{rc}}$}, \textit{l$_{\rm{rw}}$} are the large-scale path loss coefficients in the form of $\frac{\rho_0}{d^\alpha}$, where $\rho_0$ is the reference power gain at a distance of one meter ($m$), $\alpha$ indicates the path-loss exponent, and $d$ represents to the node distances of \textit{d$_{\rm{AR}}$} and \textit{d$_{\rm{rb}}$}, \textit{d$_{\rm{rc}}$}, \textit{d$_{\rm{rw}}$}.
%In addition, \textit{l$_{\rm{AR}}$} and \textit{l$_{\rm{rb}}$}, \textit{l$_{\rm{rc}}$}, \textit{l$_{\rm{rw}}$} are the large-scale path loss coefficients  corresponding to the node distances of \textit{d$_{\rm{AR}}$} and \textit{d$_{\rm{rb}}$}, \textit{d$_{\rm{rc}}$}, \textit{d$_{\rm{rw}}$}.
In this paper, we assume that the instantaneous CSI between STAR-RIS and Alice, Bob, Carol ($\mathbf{h}_{\mathrm{AR}}$, $\mathbf{h}_{\mathrm{rb}}$,  $\mathbf{h}_{\mathrm{rc}}$) is available at Alice, while only the statistical CSI between STAR-RIS and Willie ($\mathbf{h}_{\mathrm{rw}}$) is known at Alice. In contrast, %it is assumed
we consider the worst case that Willie is capable to know the global instantaneous CSI. %, i.e.,  $\mathbf{h}_{\mathrm{AR}}$, $\mathbf{h}_{\mathrm{rw}}$, $\mathbf{h}_{\mathrm{rb}}$ and $\mathbf{h}_{\mathrm{rc}}$.

When Alice communicates with Bob and Carol, % with the help of STAR-RIS,
the received signals at Bob and Carol can be respectively expressed as
\vspace{-1mm}
\begin{align}
  	& y_{\mathrm{b}}[k]=\mathbf{h}_{\mathrm{rb}}^H \boldsymbol{\Theta}_{\mathrm{r}} \mathbf{h}_{\mathrm{AR}}\big(\sqrt{P_{\mathrm{b}}} s_{\mathrm{b}}[k]+\sqrt{P_{\mathrm{c}}} s_{\mathrm{c}}[k]\big)+n_{\mathrm{b}}[k], \\
  	& y_{\mathrm{c}}[k]=\mathbf{h}_{\mathrm{rc}}^H\boldsymbol{\Theta}_{\mathrm{t}} \mathbf{h}_{\mathrm{AR}}\big(\sqrt{P_{\mathrm{b}}} s_{\mathrm{b}}[k]+\sqrt{P_{\mathrm{c}}} s_{\mathrm{c}}[k]\big)+n_{\mathrm{c}}[k],
 \end{align}
where $k \in \mathcal{K}\triangleq\{1, \ldots, K\}$ is the index of communication channel use in a time \mbox{slot}.
$\boldsymbol{\Theta}_\mathrm{\chi}=\operatorname{Diag}\Big\{\sqrt{\beta_\mathrm{\chi}^1}e^{\mathrm{j} \phi_\mathrm{\chi}^1}, \ldots,\sqrt{\beta_\mathrm{\chi}^M}$
$e^{\mathrm{j}\phi_\mathrm{\chi}^M}\Big\}$ %and
%$\boldsymbol{\Theta}_\mathrm{t}=\operatorname{Diag}\Big\{\sqrt{\beta_\mathrm{t}^1} e^{\mathrm{j} \phi_\mathrm{t}^1}, \ldots, \sqrt{\beta_\mathrm{t}^M}e^{\mathrm{j}\phi_\mathrm{t}^M}\Big\}$
with $\chi\in\{\mathrm{r},\mathrm{t}\}$ indicate the reflected and transmitted coefficient matrices of STAR-RIS, where $\beta_\mathrm{\chi}^m\in[0,1]$, $\beta_\mathrm{r}^m+\beta_\mathrm{t}^m=1$ and $\phi_\mathrm{\chi}^m\in[0,2\pi)$, for $\forall m \in \mathcal{M} \triangleq\{1,2, \ldots, M\}$.
In addition, %$P_\mathrm{b}$ and $ P_\mathrm{c}$ are the power allocation at Alice for Bob and Carol, while
$s_\mathrm{b}[k]$ and $s_\mathrm{c}[k]$ $\sim \mathcal{C N}(0, 1)$ are the covert and public signals transmitted by Alice to Bob and Carol with the power allocation $P_\mathrm{b}$ and $ P_\mathrm{c}$, respectively. % where we try to hide the transmission of $s_\mathrm{b}[k]$ from the detection of Willie.
Here, $n_\mathrm{b}[k]\sim\mathcal{C N}(0, \sigma_\mathrm{b}^2)$ and $n_\mathrm{c}[k]\sim\mathcal{C N}(0, \sigma_\mathrm{c}^2)$ are the AWGN noise received at Bob and Carol with noise power $\sigma_\mathrm{b}^2$ and $\sigma_\mathrm{c}^2$.
%\section{Analysis on STAR-RIS-Assisted Covert Communications in NOMA System}\label{analysis_DEP}
%
\vspace{-2mm}
\section{Analysis on Covert Strategy}\label{analysis_DEP}
\vspace{-0.5mm}
\subsection{CC Detection Strategy at Willie}
%\vspace{-1mm}
In this section, the detection strategy of Willie for covert communications from Alice to Bob is given in details. %In particular, Willie attempts to judge whether there exists covert transmissions based on the received signal sequence $\{y_\mathrm{w}[k]\}_{k \in \mathcal{K}}$ in a time slot.
In fact, Willie faces a binary detection hypothesis  based on the received signal sequence $\{y_\mathrm{w}[k]\}_{k \in \mathcal{K}}$, i.e., a null hypothesis $\mathcal{H}_0$ and an alternative hypothesis $\mathcal{H}_1$, respectively indicating that Alice only transmits public signals to Carol or both public and covert signals to Coral and Bob. Accordingly, the received signals at Willie for the two hypotheses are given by

\vspace{-3mm}
{\small{
\begin{align}
 	 \mathcal{H}_0: y_{\mathrm{w}}[k]=&\mathbf{h}_{\mathrm{rw}}^H  \boldsymbol{\Theta}_{\mathrm{r}} \mathbf{h}_{\mathrm{AR}} \sqrt{P_{\mathrm{c}}} s_{\mathrm{c}}[k]+n_{\mathrm{w}}[k], \\
 	 \mathcal{H}_1: y_{\mathrm{w}}[k]=&\mathbf{h}_{\mathrm{rw}}^H  \boldsymbol{\Theta}_{\mathrm{r}} \mathbf{h}_{\mathrm{AR}}\big(\sqrt{P_{\mathrm{b}}} s_{\mathrm{b}}[k]+\sqrt{P_{\mathrm{c}}} s_{\mathrm{c}}[k]\big)+n_{\mathrm{w}}[k],
\end{align}
}}
\hspace{-1.2mm}where $n_\mathrm{w}[k]\sim\mathcal{C N}(0, \sigma_\mathrm{w}^2)$ represents the background noise at Willie, which introduces the uncertainty to confuse
Willie's detection for the covert communications.
%The uncertainty of the background noise at Willie is considered to weaken its detection capability for communications between Alice and Bob \cite{he2017covert}.
The probability density function (PDF) of the noise power $\sigma_\mathrm{w}^2$ is given as \cite{he2017covert} %\cite{he2017covert}
\begin{align}\label{noise_PDF}
f_{\sigma_{\mathrm{w}}^2}(x)= \begin{cases}\frac{1}{2 x \ln \rho}, & \frac{\hat{\sigma}_{\mathrm{w}}^2}{\rho} \leq x \leq \rho \hat{\sigma}_{\mathrm{w}}^2, \\
	0, & \text { otherwise },\end{cases}
\end{align}
where $\rho > 1$ is the parameter that quantifies the size of the uncertainty, $\hat{\sigma}_{\mathrm{w}}^2$ represents the nominal noise power. %Thus, it is difficult for Willie to detect the existence of communications between Alice and Bob under the random noise.
Similar to \cite{tao20}, we assume that Willie utilizes a radiometer to \mbox{detect} covert transmissions, % due to its low complexity and ease of implementation \cite{tao20}.
%According to the working mechanism of the radiometer,
where the average power of the received signals in a time slot, i.e., $\overline{P}_\mathrm{w}=\frac{1}{K} \sum_{k=1}^K\left|y_\mathrm{w}[k]\right|^2$, is employed for statistical test. In addition, it is assumed that Willie takes infinite number of signal samples, i.e., $K \rightarrow \infty$, to implement binary detection  \cite{Hu19,Wang21}. %, zheng21
%The reasonability of this assumption can be justified as follows. First of all, as indicated in \cite{Wang21}, Willie uses the infinite number of symbols to establish an upper bound about the number of received symbols. Second, due to the communication bandwidth directly affects the number of transmitted symbols in a time slot, i.e. the wider the bandwidth is, the more symbols can be transmitted within each time slot. Therefore, we can choose an appropriate communication bandwidth to make the number of symbols transmitted in a time slot sufficiently large \cite{zhou19}.
Therefore, the average received power  $\overline{P}_\mathrm{w}$ can be asymptotically approximated as
\vspace{-1mm}
\begin{align}\label{eq_repower_w}
		\overline{P}_\mathrm{w}=\begin{cases}P_\mathrm{c}\left|\mathbf{h}_{\mathrm{rw}}^H \boldsymbol{\Theta}_\mathrm{r} \mathbf{h}_{\mathrm{AR}}\right|^2+\sigma_\mathrm{w}^2, & \mathcal{H}_0, \\\left(P_\mathrm{b}+P_\mathrm{c}\right)\left|\mathbf{h}_{\mathrm{rw}}^H \boldsymbol{\Theta}_\mathrm{r} \mathbf{h}_{\mathrm{AR}}\right|^2+\sigma_\mathrm{w}^2, & \mathcal{H}_1.\end{cases}
\end{align}
%\footnote{We ignore the noise uncertainty at Bob for two reasons. First, the noise uncertainty at Willie is crucial for CC as it prevents Willie from detecting the covert information, while at Bob, it only reduces the rate performance. Second, the noise uncertainty at Willie can be induced by authorized users who emit jamming signals to confuse Willie \cite{he2017covert}.}
%Based on $\overline{P}_\mathrm{w}$, Willie makes the decision of whether the communication between Alice and Bob is under the  hypotheses of $\mathcal{H}_0$ or $ \mathcal{H}_1$ with the decision rule of $\overline{P}_\mathrm{w} \underset{\mathcal{D}_0}{\stackrel{\mathcal{D}_1}{\gtrless}} \tau_\mathrm{dt}$, where $\mathcal{D}_0$ (or $\mathcal{D}_1$) indicates the decision that Willie favors $\mathcal{H}_0$ (or $\mathcal{H}_1$), and $\tau_\mathrm{d t}>0$ is the corresponding detection threshold.

Based on $\overline{P}_\mathrm{w}$, Willie makes the decision with the rule of
$\overline{P}_\mathrm{w} \underset{\mathcal{D}_0}{\stackrel{\mathcal{D}_1}{\gtrless}}\tau_\mathrm{dt}$, where $\mathcal{D}_0$ or $\mathcal{D}_1$  indicates the decision that Willie favors the hypotheses of $\mathcal{H}_0$ or $\mathcal{H}_1$, and $\tau_\mathrm{d t}>0$ is the detection threshold.
Willie's detection performance is measured by the detection error probability (DEP) denoted as $P_\mathrm{e}\in[0,1]$, and we consider the worst case  that  Willie can optimize its detection threshold $\tau_\mathrm{d t}$ to minimize the DEP.
%According to the Neyman-Pearson criterion, the minimum DEP of Willie is the likelihood ratio \cite{Wang21, li20}, which can be expressed as %$\Lambda\left(\mathbf{y}_w\right)=\frac{f_{\mathbf{y}_w \mid H_1}\left(\mathbf{y}_w \mid H_1\right)}{f_{\mathbf{y}_w \mid H_0}\left(\mathbf{y}_w \mid H_0\right)}$,
%\begin{align}\label{eq_likelihood_w}
%\Lambda\left(\mathbf{y}_\mathrm{w}\right)=\frac{f_{\mathbf{y}_\mathrm{w} \mid \mathcal{H}_1}\left(\mathbf{y}_\mathrm{w} \mid \mathcal{H}_1\right)}{f_{\mathbf{y}_\mathrm{w} \mid \mathcal{H}_0}\left(\mathbf{y}_\mathrm{w} \mid \mathcal{H}_0\right)},
%\end{align}
%where $\mathbf{y}_\mathrm{w}=\left\{y_\mathrm{w}[1], \ldots, y_\mathrm{w}[K]\right\}$ is Willie's received signal vector, $f_{\mathbf{y}_\mathrm{w} \mid \mathcal{H}_0}$ and $f_{\mathbf{y}_\mathrm{w} \mid \mathcal{H}_1}$ are the probability density functions (PDFs) of the sampling signals when $\mathcal{H}_0$ and $\mathcal{H}_1$ are true, respectively.
%However, it is difficult to derive the likelihood ratio due to the fact that the instantaneous CSI of $ \mathbf{h}_{\mathrm{AR}} $ is unavailable at Willie, which introduces extra randomness in received signal $\mathbf{y}_\mathrm{w}$.
It is known that the DEP $P_\mathrm{e}$ is the sum of the false alarm (FA) probability $P_{\mathrm{FA}}=\operatorname{Pr}\left(\mathcal{D}_1 \mid \mathcal{H}_0\right)$ and the miss detection (MD) probability $P_{\mathrm{MD}}=\operatorname{Pr}\left(\mathcal{D}_0 \mid \mathcal{H}_1\right)$ which respectively represent the probabilities of Willie making the decision $\mathcal{D}_1$ under $\mathcal{H}_0$ or $\mathcal{D}_0$ under $\mathcal{H}_1$. %Hence, the DEP of Willie can be written as $P_\mathrm{e}=P_{\mathrm{FA}}+P_{\mathrm{MD}}=\operatorname{Pr}\left(\mathcal{D}_0 \mid \mathcal{H}_1\right)+\operatorname{Pr}\left(\mathcal{D}_1 \mid \mathcal{H}_0\right)$.
%To determine the minimum DEP, we first derive the DEP based on the false alarm (FA) and miss detection (MD) probabilities and then use this information to obtain the minimum DEP,
%It is easy to note that $0 \leq P_\mathrm{e} \leq 1$.
%
%\subsection{ Analysis on Detection Error Probability}
%In this section, we first give the expression for $P_{\mathrm{e}}$, which can be expressed as
Hence, we can derive the DEP as
\vspace{-1mm}
\begin{align}\label{eq_DEP}
  P_\mathrm{e}=&P_\mathrm{FA}+P_\mathrm{MD}\notag\\
  =&\operatorname{Pr}\left(\overline{P}_\mathrm{w}>\tau_\mathrm{dt}|\mathcal{H}_0\right)+\operatorname{Pr}\left(\overline{P}_\mathrm{w}<\tau_\mathrm{dt}|\mathcal{H}_1\right)\notag\\
  =&\operatorname{Pr}\left(\sigma_\mathrm{w}^2>\tau_\mathrm{dt}-\phi_1\right)+\operatorname{Pr}\left(\sigma_\mathrm{w}^2<\tau_\mathrm{dt}-\phi_2\right)\notag\\
  =&1-\operatorname{Pr}\left(\tau_\mathrm{dt}-\phi_2\leq\sigma_\mathrm{w}^2\leq\tau_\mathrm{dt}-\phi_1\right),
\end{align}
where $\phi_1= P_\mathrm{c}\left|\mathbf{h}_{\mathrm{rw}}^H \boldsymbol{\Theta}_\mathrm{r} \mathbf{h}_{\mathrm{AR}}\right|^2$, $\phi_2=\left(P_\mathrm{b}+P_\mathrm{c}\right)$$\left|\mathbf{h}_{\mathrm{rw}}^H \boldsymbol{\Theta}_\mathrm{r} \mathbf{h}_{\mathrm{AR}}\right|^2$. It is assumed that the detection threshold $\tau_\mathrm{dt}\in\big[\frac{\hat{\sigma}_{\mathrm{w}}^2}{\rho}+\phi_1,$ $\rho\hat{\sigma}_{\mathrm{w}}^2+\phi_1\big]$ so that the uncertainty of the background noise at
Willie can be fully used to explore its effect on the detection ability. Based on this assumption and the PDF of noise power at Willie, the analytical expression of $ P_\mathrm{e}$ is given by
\vspace{-1mm}
\begin{align}\label{eq_DEP_expression}
 P_\mathrm{e}=&1-\int_{\max\left\{\tau_\mathrm{dt}-\phi_2, \frac{\hat{\sigma}_\mathrm{w}^2}{\rho}\right\}}^{\tau_\mathrm{dt}-\phi_1}\frac{1}{2x\ln\rho}dx\notag=1-\frac{1}{2\ln\rho}\times\\
 &\left(\ln\left(\tau_\mathrm{dt}-\phi_1\right)-\ln\left(\max\Big\{\tau_\mathrm{dt}-\phi_2, \frac{\hat{\sigma}_\mathrm{w}^2}{\rho}\Big\}\right)\right).
\end{align}

%It should be noted that we assume a worst-case scenario where Willie can optimize its detection threshold to minimize the DEP.
%By choosing a reasonable
%By optimizing the detection threshold $\tau_\mathrm{d t}$, the minimum DEP, denoted as $P_\mathrm{e}^*$, can be obtained at Willie.
%The following theorem shows the closed-form solution of the optimal $\tau_\mathrm{dt}^*$.
\vspace{-2mm}
\begin{theorem}
The closed-form optimal detection threshold $\tau_\mathrm{dt}^*$ to minimize the DEP at Willie in the considered STAR-RIS-assisted covert NOMA communication system is given as	
\begin{align}\label{eq_minimum_threshold}
     \tau_\mathrm{dt}^*=\min\left\{\phi_2+ \frac{\hat{\sigma}_\mathrm{w}^2}{\rho}, \phi_1+\rho\hat{\sigma}_\mathrm{w}^2\right\},
\end{align}
\begin{proof}
	The proof is given in Appendix \ref{Apd_A}.
\end{proof}
\end{theorem}

Substituting optimal detection threshold $\tau_\mathrm{dt}^*$  into \eqref{eq_DEP_expression} and adopting some algebraic manipulations, the analytical closed-form expression of the minimum DEP can be derived as
\vspace{-2mm}
\begin{align}
%P_\mathrm{e}^*=\begin{cases}
% 1-\frac{\ln\left(1+\frac{\rho\left(\phi_2-\phi_2\right)}{\hat{\sigma}_\mathrm{w}^2}\right)}{2\ln\rho}, &\phi_2-\phi_1\leq \rho\hat{\sigma}_\mathrm{w}^2-\frac{\hat{\sigma}_\mathrm{w}^2}{\rho},\\
%0, &\mathrm{otherwise}.
P_\mathrm{e}^*=\begin{cases}
 1-\frac{\ln\left(1+\frac{\rho\left(\phi_2-\phi_1\right)}{\hat{\sigma}_\mathrm{w}^2}\right)}{2\ln\rho}, &\phi_2-\phi_1\leq \frac{(\rho^2-1)\hat{\sigma}_\mathrm{w}^2}{\rho},\\
0, &\mathrm{otherwise}.
\end{cases}
\end{align}
In this paper,  $\phi\triangleq \phi_2-\phi_1=P_\mathrm{b}\left|\mathbf{h}_{\mathrm{rw}}^H \boldsymbol{\Theta}_\mathrm{r} \mathbf{h}_{\mathrm{AR}}\right|^2\leq \frac{(\rho^2-1)\hat{\sigma}_\mathrm{w}^2}{\rho}$.
To guarantee the covertness of communications between Alice and Bob, $P_\mathrm{e}^*\geq 1-\epsilon$ is required, where $\epsilon \in(0,1)$ is determined  by the system performance indicators. Based
on this requirement, we can further obtain the constraint
$\phi\leq \min\Big\{\frac{\left(\rho^{2\epsilon}-1\right)\hat{\sigma}_\mathrm{w}^2}{\rho}, \frac{(\rho^2-1)\hat{\sigma}_\mathrm{w}^2}{\rho}\Big\}=\frac{\left(\rho^{2\epsilon}-1\right)\hat{\sigma}_\mathrm{w}^2}{\rho}$  to characterize the covert performance. Considering that Alice only possesses the statistical CSI of $\mathbf{h}_{\mathrm{rw}}$, the average of $\phi$ over $\mathbf{h}_{\mathrm{rw}}$, denoted as $\overline{\phi}=\mathbb{E}(\phi)_{\mathbf{h}_{\mathrm{rw}}}$, is utilized to evaluate the covert communications between Alice and Bob. $\overline{\phi}$
can be expressed as $\overline{\phi}=\int_{	0}^{+\infty}x\frac{e^{-\frac{x}{\lambda}}}{\lambda}dx=\lambda$,
where $\lambda=P_\mathrm{b}l_\mathrm{rw}\left\|\boldsymbol{\Theta}_\mathrm{r} \mathbf{h}_{\mathrm{AR}}\right\|^2$. Therefore, the covert constraint
can be given as %$\lambda\leq\min\Big\{\frac{\left(\rho^{2\epsilon}-1\right)\hat{\sigma}_\mathrm{w}^2}{\rho}, \rho\hat{\sigma}_\mathrm{w}^2-\frac{\hat{\sigma}_\mathrm{w}^2}{\rho}\Big\}$. Considering that $\epsilon\in(0, 1)$, the the covert constraint can be written as
$\lambda\leq \frac{1}{\rho}\left(\rho^{2\epsilon}-1\right)\hat{\sigma}_\mathrm{w}^2$.

\vspace{-2mm}
\subsection{Transmissions between Alice and Legitimate Users}
In this paper, the legitimate users' CSI is adopted to determine the successive interference cancelation (SIC) decoding order in NOMA systems, which is a straightforward way being widely used in existing literature \cite{tao20, lv2021covert}. It is assumed that Carol is allocated with more transmit power than Bob, i.e., $P_\mathrm{c}\geq P_\mathrm{b}$, so that the higher power of Carol can help to hide the covert transmissions between Alice and Bob from the detection of Willie in practical applications.
%With this assumption, \textcolor[rgb]{1.00,0.00,0.00}{Bob is required to hold a higher channel gain than Carol to ensure a successful SIC}.
%\textcolor[rgb]{1.00,0.00,0.00}{According to this SIC decoding order},
Hence, Bob first decodes $s_\mathrm{c}[k]$ and eliminates it from the received signals, and then decodes its own signal $s_\mathrm{b}[k]$. Hence, the available decoding rates at Bob for $s_\mathrm{b}$ and $s_\mathrm{c}$  are respectively expressed as

\vspace{-3mm}
{\small{
\begin{align}
    &R_\mathrm{bc}=\log_2\bigg(1+\frac{P_\mathrm{c}\left|\mathbf{h}_\mathrm{rb}^H\boldsymbol{\Theta}_\mathrm{r}\mathbf{h}_\mathrm{AR}|\right.^2}
    {P_\mathrm{b}\left|\mathbf{h}_\mathrm{rb}^H\boldsymbol{\Theta}_\mathrm{r}\mathbf{h}_\mathrm{AR}|\right.^2
   +\sigma_\mathrm{b}^2}\bigg),\\
   &R_\mathrm{bb}=\log_2\bigg(1+\frac{P_\mathrm{b}\left|\mathbf{h}_\mathrm{rb}^H\boldsymbol{\Theta}_\mathrm{r}\mathbf{h}_\mathrm{AR}|\right.^2}
    {\sigma_\mathrm{b}^2}\bigg).
\end{align}
}}
For Carol, $s_\mathrm{c}$ is directly decoded by treating $s_\mathrm{b}$ as interference, and the available rate is given by

\vspace{-3mm}
{\small{
\begin{align}
    R_\mathrm{cc}=\log_2\bigg(1+\frac{P_\mathrm{c}\left|\mathbf{h}_\mathrm{rc}^H\boldsymbol{\Theta}_\mathrm{t}\mathbf{h}_\mathrm{AR}|\right.^2}
    {P_\mathrm{b}\left|\mathbf{h}_\mathrm{rc}^H\boldsymbol{\Theta}_\mathrm{t}\mathbf{h}_\mathrm{AR}\right|^2
    +\sigma_\mathrm{c}^2}\bigg).
\end{align}
}}

\vspace{-4mm}
\section{Problem Formulation and Algorithm Design}\label{P_A}
%\vspace{-1mm}
\subsection{Optimization Problem Formulation}
\vspace{-1mm}
%Based on the discussions in section \ref{analysis_DEP},
This section formulates an optimization problem to maximize the covert rate between Alice and Bob while ensuring the QoS at Carol, which is given below
\vspace{-1mm}
\begin{subequations}\label{eq_ori_opt}
  \begin{align}
    &\max _{P_\mathrm{b}, P_\mathrm{c}, \boldsymbol{\Theta}_\mathrm{r}, \boldsymbol{\Theta}_\mathrm{t}} ~~R_\mathrm{bb},\notag \\
    &\quad~\text { s.t. }  P_\mathrm{b}+P_\mathrm{c}\leq P_\mathrm{tmax}, P_\mathrm{c}\geq P_\mathrm{b},\label{eq_ori_opt_1}\\
    &\qquad~~~~R_\mathrm{bc}\geq R_\mathrm{cc},\label{eq_ori_opt_2}\\
    &\qquad~~~~R_\mathrm{cc}\geq R^*,\label{eq_ori_opt_3}\\
    &\qquad~~~~\lambda\leq \frac{1}{\rho}\left(\rho^{2\epsilon}-1\right)\hat{\sigma}_\mathrm{w}^2,\label{eq_ori_opt_4}\\
    &\qquad~~~~\beta_\mathrm{r}^m+\beta_\mathrm{t}^m=1, \phi_\mathrm{r}^m, \phi_\mathrm{t}^m\in[0, 2\pi), m\in\mathcal{M},\label{eq_ori_opt_5}
  \end{align}
\end{subequations}
through jointly optimizing Alice's transmit power and STAR-RIS's passive beamforming variables, i.e., $P_\mathrm{b}$, $P_\mathrm{c}$, $\boldsymbol{\Theta}_\mathrm{r}$ and $\boldsymbol{\Theta}_\mathrm{t}$. %Hence, the optimization problem is formulated as
Here, \eqref{eq_ori_opt_1} includes the transmit power constraints  with $P_\mathrm{tmax}$ being the maximum transmit power of Alice; % and the power allocation constraint for Bob and Carol;
\eqref{eq_ori_opt_2} is the constraint that guarantees the successful SIC at Bob;  \eqref{eq_ori_opt_3} represents the QoS constraint for Carol;  \eqref{eq_ori_opt_4} is an equivalent covert communication constraint of $P_\mathrm{e}^*\geq 1-\epsilon$;  \eqref{eq_ori_opt_5} shows the amplitude and phase shift constraints for STAR-RIS. Actually, it is challenging to solve the formulated optimization problem because of  the strong coupling among the optimization variables.
%the following reasons. Firstly, the transmit power and passive beamforming variables, $P_\mathrm{b}$, $P_\mathrm{c}$, $\boldsymbol{\Theta}_\mathrm{r}$ and $\boldsymbol{\Theta}_\mathrm{t}$ are strongly coupled in the objective function, covert communication constraint \eqref{eq_ori_opt_4} and QoS constraint \eqref{eq_ori_opt_3}. In addition, the utilization of STAR-RIS introduces a
%characteristic amplitude constraint \eqref{eq_ori_opt_5} due to the fact that $\boldsymbol{\Theta}_\mathrm{r}$ and $\boldsymbol{\Theta}_\mathrm{t}$ depend on each other in terms of element amplitudes.
To tackle this issue, the alternating strategy is leveraged to design the optimization algorithm.
Specifically, we divide the original problem into two subproblems where one subproblem is focused on  power allocation for Bob and Carol, i.e., $P_\mathrm{b}$, $P_\mathrm{c}$, while the other subproblem designs the passive beamformer variables $\boldsymbol{\Theta}_\mathrm{r}$ and $\boldsymbol{\Theta}_\mathrm{t}$.
\vspace{-2mm}
\subsection{Power Allocation Design}
In this section, the transmit power allocation for Bob and Carol are obtained by solving the original optimization problem \eqref{eq_ori_opt} with the given passive beamforming variables $\boldsymbol{\Theta}_\mathrm{r}$ and $\boldsymbol{\Theta}_\mathrm{t}$. Specifically, %with the given $\boldsymbol{\Theta}_\mathrm{r}$ and $\boldsymbol{\Theta}_\mathrm{t}$,
the objective function turns into maximizing  $P_\mathrm{b}$, and the corresponding subproblem can be formulated as
%\begin{subequations}\label{eq_power_allocation}
%  \begin{align}
%    &\max _{P_\mathrm{b}, P_\mathrm{c}} ~~ P_\mathrm{b},\notag \\
%    &\quad~\text { s.t. }  P_\mathrm{b}+P_\mathrm{c}\leq P_\mathrm{tmax}, P_\mathrm{c}\geq P_\mathrm{b},\label{eq_power_allocation_1}\\
%    &\qquad~~~~R_\mathrm{bc}\geq R_\mathrm{cc},\label{eq_power_allocation_2}\\
%    &\qquad~~~~R_\mathrm{cc}\geq R^*,\label{eq_power_allocation_3}\\
%    &\qquad~~~~\lambda\leq \frac{1}{\rho}\left(\rho^{2\epsilon}-1\right)\hat{\sigma}_\mathrm{w}^2.\label{eq_power_allocation_4}
% \end{align}
%\end{subequations}
\vspace{-1mm}
\begin{align}
    &\max _{P_\mathrm{b}, P_\mathrm{c}} ~~ P_\mathrm{b},\notag \\
    &\quad\text { s.t. } \eqref{eq_ori_opt_1}-\eqref{eq_ori_opt_4}.\label{eq_power_allocation}
\end{align}

%We first handle the constraint \eqref{eq_power_allocation_2} by simplifying it, we can obtain $\sigma_{\mathrm{c}}^2\left|\mathbf{h}_\mathrm{rb}^H\boldsymbol{\Theta}_\mathrm{r}\mathbf{h}_\mathrm{AR}\right|^2\geq\sigma_{\mathrm{b}}^2\left|\mathbf{h}_\mathrm{rc}^H\boldsymbol{\Theta}_\mathrm{t}\mathbf{h}_\mathrm{AR}\right|^2$, which is irrelevant with the $P_\mathrm{b}$ and $P_\mathrm{c}$. Thus, the the constraint \eqref{eq_power_allocation_2} can be removed from optimization problem \eqref{eq_power_allocation}. Similarily, by equivalently transforming the QoS constraint \eqref{eq_power_allocation_3} and the covert constraint \eqref{eq_power_allocation_4}, the optimization problem \eqref{eq_power_allocation} can be simplified as
We first deal with the constraint \eqref{eq_ori_opt_2} and can obtain that $\sigma_{\mathrm{c}}^2\left|\mathbf{h}_\mathrm{rb}^H\boldsymbol{\Theta}_\mathrm{r}\mathbf{h}_\mathrm{AR}\right|^2\geq\sigma_{\mathrm{b}}^2\left|\mathbf{h}_\mathrm{rc}^H\boldsymbol{\Theta}_\mathrm{t}\mathbf{h}_\mathrm{AR}\right|^2$, which is irrelevant with $P_\mathrm{b}$ and $P_\mathrm{c}$. Thus, constraint \eqref{eq_ori_opt_2} can be removed from the  problem \eqref{eq_power_allocation}. Similarly, by equivalently transforming the QoS constraint \eqref{eq_ori_opt_3} and the covert constraint \eqref{eq_ori_opt_4}, the optimization problem \eqref{eq_power_allocation} can be simplified as

\vspace{-4mm}
{\small{
\begin{subequations}\label{eq_power_allocation1}
  \begin{align}
    &\hspace{-4mm}\quad~\max _{P_\mathrm{b}, P_\mathrm{c}} ~~P_\mathrm{b},\notag \\
    &\hspace{-4mm}\quad~~\text {s.t.}~~\eqref{eq_ori_opt_1},  \label{eq_power_allocation1_1}\\
    &\hspace{-4mm}\qquad~~~~P_\mathrm{b}\leq\frac{P_\mathrm{c}\left|\mathbf{h}_\mathrm{rc}^H\boldsymbol{\Theta}_\mathrm{t}\mathbf{h}_\mathrm{AR}\right|^2}{\left(2^{R^*}-1\right)\left|\mathbf{h}_\mathrm{rc}^H\boldsymbol{\Theta}_\mathrm{t}\mathbf{h}_\mathrm{AR}\right|^2}-\frac{\sigma_{\mathrm{c}}^2}{\left|\mathbf{h}_\mathrm{rc}^H\boldsymbol{\Theta}_\mathrm{t}\mathbf{h}_\mathrm{AR}\right|^2}, \label{eq_power_allocation1_2}\\
    &\hspace{-4mm}\qquad~~~~P_\mathrm{b}\leq\frac{\left(\rho^{2\epsilon}-1\right)\hat{\sigma}_\mathrm{w}^2}{\rho l_\mathrm{rw}\left\|\boldsymbol{\Theta}_\mathrm{r} \mathbf{h}_{\mathrm{AR}}\right\|^2}, \label{eq_power_allocation1_3}
    \end{align}
\end{subequations}
}}
%As a result, the problem \eqref{eq_power_allocation1}
\hspace{-1.2mm}which is a simple linear programming problem. It is easy to derive the optimal solution as $P_\mathrm{b}^*=\min\Big\{\frac{P_\mathrm{tmax}}{2}, \frac{\left(\rho^{2\epsilon}-1\right)\hat{\sigma}_\mathrm{w}^2}{\rho l_\mathrm{rw}\left\|\boldsymbol{\Theta}_\mathrm{r} \mathbf{h}_{\mathrm{AR}}\right\|^2},$ $ \frac{P_\mathrm{tmax}\left|\mathbf{h}_\mathrm{rc}^H\boldsymbol{\Theta}_\mathrm{t}\mathbf{h}_\mathrm{AR}\right|^2-\left(2^{R^*}-1\right)
\sigma_{\mathrm{c}}^2}{2^{R^*}\left|\mathbf{h}_\mathrm{rc}^H\boldsymbol{\Theta}_\mathrm{t}\mathbf{h}_\mathrm{AR}\right|^2}\Big\}$ and $P_\mathrm{c}^*=P_\mathrm{tmax}-P_\mathrm{b}^*$.
\vspace{-1mm}
\subsection{Joint Passive Beamforming Design for STAR-RIS}
\vspace{-1mm}
%After obtaining the transmit power allocation coefficients,
In this subsection, the passive reflecting and transmitting beamforming variables $\boldsymbol{\Theta}_\mathrm{r}$ and $\boldsymbol{\Theta}_\mathrm{t}$ of STAR-RIS are jointly designed by solving the optimization problem \eqref{eq_ori_opt} with the acquired $P_\mathrm{b}$ and $P_\mathrm{c}$ in previous subsection. %In particular, on the basis of the original problem \eqref{eq_ori_opt},
The corresponding subproblem  can be expressed as %for joint reflect and transmit beamforming design of STAR-RIS
\vspace{-2mm}
%\begin{subequations}\label{eq_passive_opt}
%  \begin{align}
%    &\max _{\boldsymbol{\Theta}_\mathrm{r}, \boldsymbol{\Theta}_\mathrm{t}} ~~R_\mathrm{bb},\notag \\
%    &\quad\text { s.t. } ~\eqref{eq_ori_opt_2}, \eqref{eq_ori_opt_3}, \eqref{eq_ori_opt_4}, \eqref{eq_ori_opt_5}. \label{eq_passive_opt_1}
 %  \end{align}
%\end{subequations}
\begin{align}
    &\max _{\boldsymbol{\Theta}_\mathrm{r}, \boldsymbol{\Theta}_\mathrm{t}} ~~R_\mathrm{bb},\notag \\
    &\quad\text { s.t. } ~\eqref{eq_ori_opt_2}-\eqref{eq_ori_opt_5}. \label{eq_passive_opt} %\label{eq_passive_opt_1}
\end{align}
Note that constraint \eqref{eq_ori_opt_2} and \eqref{eq_ori_opt_3} are non-convex with respect to (w.r.t.) $\boldsymbol{\Theta}_\mathrm{r}$ and $\boldsymbol{\Theta}_\mathrm{t}$, which makes this problem difficult to be solved directly. To effectively tackle this issue, we resort to the semi-definite relaxation (SDR) method. Specifically, let %$\mathbf{Q}_\mathrm{r}=\boldsymbol{\vartheta}_\mathrm{r}^*\boldsymbol{\vartheta}_\mathrm{r}^T$, $\mathbf{Q}_\mathrm{t}=\boldsymbol{\vartheta}_\mathrm{t}^*\boldsymbol{\vartheta}_\mathrm{t}^T$,
$\mathbf{Q}_\mathrm{\chi}=\boldsymbol{\vartheta}_\mathrm{\chi}^*\boldsymbol{\vartheta}_\mathrm{\chi}^T$,
$\boldsymbol{\beta}_\mathrm{\chi}=\left[\beta_\mathrm{\chi}^1, \cdots, \beta_\mathrm{\chi}^M\right]$,
%$\boldsymbol{\beta}_\mathrm{r}=\left[\beta_\mathrm{r}^1, \cdots, \beta_\mathrm{r}^M\right]$, $\boldsymbol{\beta}_\mathrm{t}=\left[\beta_\mathrm{t}^1, \cdots, \beta_\mathrm{t}^M\right]$,
where $\boldsymbol{\vartheta}_\mathrm{\chi}=\operatorname{diag}(\boldsymbol{\Theta}_\mathrm{\chi})$ with $\chi\in\{\mathrm{r},\mathrm{t}\}$.
%where $\boldsymbol{\vartheta}_\mathrm{r}=\operatorname{diag}(\boldsymbol{\Theta}_\mathrm{r})$, $\boldsymbol{\vartheta}_\mathrm{t}=\operatorname{diag}(\boldsymbol{\Theta}_\mathrm{t})$.
It is easy to verify that $\mathbf{h}_\mathrm{rb}^H\boldsymbol{\Theta}_\mathrm{r}\mathbf{h}_\mathrm{AR}=\boldsymbol{\vartheta}_\mathrm{r}^T\mathbf{H}_\mathrm{rb}^*\mathbf{h}_\mathrm{AR}$, $\mathbf{h}_\mathrm{rc}^H\boldsymbol{\Theta}_\mathrm{t}\mathbf{h}_\mathrm{AR}=\boldsymbol{\vartheta}_\mathrm{t}^T\mathbf{H}_\mathrm{rc}^*\mathbf{h}_\mathrm{AR}$ and $\left\|\boldsymbol{\Theta}_\mathrm{r}\mathbf{h}_\mathrm{AR}\right\|_2^2=\boldsymbol{\beta}_\mathrm{r}^T\left(\mathbf{h}_\mathrm{AR}\circ \mathbf{h}_\mathrm{AR}^*\right)$, where $\mathbf{H}_\mathrm{rb}=\operatorname{Diag}(\mathbf{h}_\mathrm{rb})$, $\mathbf{H}_\mathrm{rc}=\operatorname{Diag}(\mathbf{h}_\mathrm{rc})$ and $\circ$ means Hadamard product. Hence, problem \eqref{eq_passive_opt} can be equivalently transformed as

\vspace{-4mm}
{\small{
\begin{subequations}\label{eq_passive_opt_trans}
  \begin{align}
    &\hspace{-2mm}\max _{\mathbf{Q}_\mathrm{r}, \mathbf{Q}_\mathrm{t}, \boldsymbol{\beta}_\mathrm{r}, \boldsymbol{\beta}_\mathrm{t}} \operatorname{Tr}(\mathbf{Q}_\mathrm{r}\mathbf{A}) ,\notag \\
    &\hspace{-3mm}\quad~\text { s.t. } \sigma_{\mathrm{c}}^2\operatorname{Tr}(\mathbf{Q}_\mathrm{r}\mathbf{A})\geq \sigma_{\mathrm{b}}^2\operatorname{Tr}(\mathbf{Q}_\mathrm{t}\mathbf{B}) ,\label{eq_passive_opt_trans_1}\\
    &\hspace{-3mm}\qquad~~~~P_\mathrm{c}\operatorname{Tr}(\mathbf{Q}_\mathrm{t}\mathbf{B})\geq \big(2^{R^*}-1\big)\big(P_\mathrm{b}\operatorname{Tr}(\mathbf{Q}_\mathrm{t}\mathbf{B})+\sigma_{\mathrm{c}}^2\big),\label{eq_passive_opt_trans_2}\\
    &\hspace{-3mm}\qquad~~~~\boldsymbol{\beta}_\mathrm{r}^T\left(\mathbf{h}_\mathrm{AR}\circ \mathbf{h}_\mathrm{AR}^*\right)\leq \frac{\left(\rho^{2\epsilon}-1\right)\hat{\sigma}_\mathrm{w}^2}{P_\mathrm{b}l_\mathrm{rw}\rho},\label{eq_passive_opt_trans_3}\\
    &\hspace{-3mm}\qquad~~~~\operatorname{diag}(\mathbf{Q}_\mathrm{r})=\boldsymbol{\beta}_\mathrm{r}, \operatorname{diag}(\mathbf{Q}_\mathrm{t})=\boldsymbol{\beta}_\mathrm{t}, \label{eq_passive_opt_trans_4}\\
    &\hspace{-3mm}\qquad~~~~\boldsymbol{\beta}_\mathrm{r}+\boldsymbol{\beta}_\mathrm{t}=\mathbf{I}_{M\times1},\label{eq_passive_opt_trans_5}\\
    &\hspace{-3mm}\qquad~~~~\mathbf{Q}_\mathrm{r}\succeq0, \mathbf{Q}_\mathrm{t}\succeq0,\label{eq_passive_opt_trans_6}\\
    &\hspace{-3mm}\qquad~~~~ \operatorname{rank}(\mathbf{Q}_\mathrm{r})=1, \operatorname{rank}(\mathbf{Q}_\mathrm{t})=1,\label{eq_passive_opt_trans_7}
  \end{align}
\end{subequations}
}}
\hspace{-1.5mm}where $\mathbf{A}=\mathbf{H}_\mathrm{rb}^*\mathbf{h}_\mathrm{AR}\mathbf{h}_\mathrm{AR}^H\mathbf{H}_\mathrm{rb}^T$, $ \mathbf{B}=\mathbf{H}_\mathrm{rc}^*\mathbf{h}_\mathrm{AR}\mathbf{h}_\mathrm{AR}^H\mathbf{H}_\mathrm{rc}^T$,
%    $\mathbf{C}=\mathbf{H}_\mathrm{AR}^H\mathbf{H}_\mathrm{AR}$.
However, problem \eqref{eq_passive_opt_trans} is still  non-convex due to the two rank-one constraints in \eqref{eq_passive_opt_trans_7}. %Considering the dependence of $\boldsymbol{\Theta}_\mathrm{r}$ and $\boldsymbol{\Theta}_\mathrm{t}$, it is difficult to re-construct the rank-one solutions if we remove the rank-one constraints directly. To handle this issue,
To solve this, we equivalently rewrite the rank-one constraints as \cite{X.HU_TCOM21RIS}
\vspace{-1mm}\begin{align}
	\eta_\mathrm{\chi}&\triangleq \operatorname{Tr}(\mathbf{Q}_\mathrm{\chi})-\left\|\mathbf{Q}_\mathrm{\chi}\right\|_2, ~\chi\in\{\mathrm{r},\mathrm{t}\},%\\
	%\eta_\mathrm{t}&\triangleq \operatorname{Tr}(\mathbf{Q}_\mathrm{t})-\left\|\mathbf{Q}_\mathrm{t}\right\|_2,
\end{align}
where $\left\|\mathbf{Q}_\mathrm{\chi}\right\|_2$ represents the spectral norm which is a convex function of $\mathbf{Q}_\mathrm{\chi}$. Note that for any positive semi-definite matrix $\mathbf{Q}\succeq0$, the inequality $\operatorname{Tr}(\mathbf{Q})-\left\|\mathbf{Q}\right\|_2\geq 0$ always holds and the equality is satisfied if and only if $\operatorname{rank}(\mathbf{Q})=1$. Based on the non-negative feature of $\eta_\mathrm{r}$ and $\eta_\mathrm{t}$, we add them into the objective function of problem \eqref{eq_passive_opt_trans} as penalty terms for the rank-one constraints. %However,  $\eta_\mathrm{r}$ and $\eta_\mathrm{t}$ are still non-convex, to address the non-convexity of  $\eta_\mathrm{r}$ and $\eta_\mathrm{t}$,
By replacing the convex spectral norms in  $\eta_\mathrm{r}$ and $\eta_\mathrm{t}$ with their linear lower-bound, i.e., first-order Taylor expansions, we can obtain the upper-bound linear approximations for $\eta_\mathrm{r}$ and $\eta_\mathrm{t}$ as

 \vspace{-4mm}
 {\small{
 \begin{align}
	\eta_\mathrm{\chi}\leq &\operatorname{Tr}(\mathbf{Q}_\mathrm{\chi})-\left(\|\mathbf{Q}_\mathrm{\chi}^{(i)}\|_2
	+\operatorname{Tr}\left(\mathbf{q}_\mathrm{\chi}^{(i)}(\mathbf{q}_\mathrm{\chi}^{(i)})^H(\mathbf{Q}_\mathrm{\chi}-\mathbf{Q}_\mathrm{\chi}^{(i)})\right)\right) \nonumber\\
	=&\widehat{\eta}_\mathrm{\chi}^{(i)}(\mathbf{Q}_\mathrm{\chi}), ~\chi\in\{\mathrm{r},\mathrm{t}\}, \label{eq_penal_trans_r} %\\
	%\eta_\mathrm{t}\leq &\operatorname{Tr}(\mathbf{Q}_\mathrm{t})-\left(\|\mathbf{Q}_\mathrm{t}^{(i)}\|_2
	%+\operatorname{Tr}\left(\mathbf{q}_\mathrm{t}^{(i)}(\mathbf{q}_\mathrm{t}^{(i)})^H(\mathbf{Q}_\mathrm{t}-\mathbf{Q}_\mathrm{t}^{(i)})\right)\right) \nonumber\\
	%=&\widehat{\eta}_\mathrm{t}^{(i)}(\mathbf{Q}_\mathrm{t}),\label{eq_penal_trans_t}
\end{align}
}}
\hspace{-2mm}where $\mathbf{q}_\mathrm{\chi}^{(i)}$ is the eigenvector corresponding to the largest eigenvalues of $\mathbf{Q}_\mathrm{\chi}^{(i)}$  in $i$-th inner loop iteration. Thus, the optimization problem \eqref{eq_passive_opt_trans} can be re-expressed as
%\begin{subequations}\label{eq_passive_opt_trans1}
%	\begin{align}
%		&\max _{\mathbf{Q}_\mathrm{r}, \mathbf{Q}_\mathrm{t}, \boldsymbol{\beta}_\mathrm{r}, \boldsymbol{\beta}_\mathrm{t}} \operatorname{Tr}(\mathbf{Q}_\mathrm{r}\mathbf{A})-\xi_1\widehat{\eta}_\mathrm{r}^{(i)}-\xi_2\widehat{\eta}_\mathrm{t}^{(i)} ,\notag \\
%		&\quad~\text { s.t. }\eqref{eq_passive_opt_trans_1}, \eqref{eq_passive_opt_trans_2}, \eqref{eq_passive_opt_trans_3}, \eqref{eq_passive_opt_trans_4}, \eqref{eq_passive_opt_trans_5}, \eqref{eq_passive_opt_trans_6},
%	\end{align}
%\end{subequations}
\begin{align}
		&\max _{\mathbf{Q}_\mathrm{r}, \mathbf{Q}_\mathrm{t}, \boldsymbol{\beta}_\mathrm{r}, \boldsymbol{\beta}_\mathrm{t}} \operatorname{Tr}(\mathbf{Q}_\mathrm{r}\mathbf{A})-\xi_1\widehat{\eta}_\mathrm{r}^{(i)}-\xi_2\widehat{\eta}_\mathrm{t}^{(i)} ,\notag \\
		&\quad~~\text { s.t. }~~~\eqref{eq_passive_opt_trans_1}-\eqref{eq_passive_opt_trans_6}, \label{eq_passive_opt_trans1}
\end{align}
where $\xi_1>0$ and $\xi_2>0$ are the introduced penalty coefficients, which will iteratively increase to obtain the solutions that meet the rank-one constraint. As a result, optimization problem \eqref{eq_passive_opt_trans1} is convex and can be effectively solved by existing convex optimization tools such as CVX \cite{grant14cvx}.
\vspace{-3mm}
\subsection{Proposed Optimization Algorithm}
\vspace{-2mm}
\begin{center}
	\begin{tabular}{p{8.5cm}}
		\toprule[2pt]
		{\small\textbf{Algorithm 1:} Proposed Algorithm for Problem \eqref{eq_ori_opt}   }\\ %solving optimized problem \eqref{eq50}
		%\textbf{Algorithm 1:} Proposed Alternating Algorithm for STAR-RIS-assisted Covert Communications Problem \eqref{eq_ori_opt}   \\ %solving optimized problem \eqref{eq50}
		\midrule[1pt]
		1:{\small~Initialize feasible point $\left(P_\mathrm{b}^{(0)}, P_\mathrm{c}^{(0)}, \boldsymbol{\Theta}_\mathrm{r}^{(0,0)}, \boldsymbol{\Theta}_\mathrm{t}^{(0,0)}\right)$; }Define\\~~~{\small the tolerance accuracy $\varepsilon$ and $\widehat{\varepsilon}$; Set the outer iteration index}
		\\~~~{\small$m$ = 0.}\\
		%~~$i=0$ for  inner loop.  \\
		2: {\small\textbf{While} $v>\varepsilon$  or $m=0$ \textbf{do}}                       \\
		3: \quad {\small Solve the linear programming problem \eqref{eq_power_allocation} and update} \\\quad~~~{\small$\Big(P_\mathrm{b}^{(m+1)}, P_\mathrm{c}^{(m+1)}\Big)$ with the obtained solutions.}\\
		4: \quad {\small Set inner iteration index $i$ = 0; Initialize $\xi_1^{(0)}$ and $\xi_2^{(0)}$.}\\
		5: \quad{\small\textbf{While} $\widehat{v}>\widehat{\varepsilon} $ or $i=0$ \textbf{do}}\\
		6: \qquad {\small Solve problem \eqref{eq_passive_opt_trans1} with given ($\boldsymbol{\Theta}_\mathrm{r}^{(m, i)}$, $\boldsymbol{\Theta}_\mathrm{t}^{(m, i)}$).} \\
		7: \qquad {\small Update the ($\boldsymbol{\Theta}_\mathrm{r}^{(m, i+1)}$, $\boldsymbol{\Theta}_\mathrm{t}^{(m, i+1)}$) with the solution.}\\
		8:  \qquad {\small Calculate $\widehat{v}=\max\{\eta_\mathrm{r}, \eta_\mathrm{t}\}$ based on the acquired solu-}\\ \quad\qquad {\small~tion; Update the penalty coefficients $\xi_1^{(i+1)}=\omega\xi_1^{(i)}$, }\\ \quad\qquad {\small ~$\xi_2^{(i+1)}=\omega\xi_2^{(i)}$ and let $i=i+1$.}\\
		9: \quad{\small\textbf{end while}}\\
		10: \quad{\small Update $\left(\boldsymbol{\Theta}_\mathrm{r}^{(m+1, 0)}, \boldsymbol{\Theta}_\mathrm{t}^{(m+1, 0)}\right)$  with $\Big(\boldsymbol{\Theta}_\mathrm{r}^{(m, i)}, \boldsymbol{\Theta}_\mathrm{t}^{(m, i)}\Big)$.}\\
		11:\quad{\small~Calculate the objective value $R_\mathrm{bb}^{(m+1)}$ and update $v=$}\\\quad\quad~{\small$\left|R_\mathrm{bb}^{(m+1)}-R_\mathrm{bb}^{(m)}\right|$; Let $m=$$m+1$.}\\
		12: {\small\textbf{end while}}      \\
		\bottomrule[2pt]
	\end{tabular}
\end{center}
%\subsection{Proposed Optimization Algorithm Analysis on Complexity }

The proposed iterative algorithm for effectively solving the initial optimization problem \eqref{eq_ori_opt} is completed by Algorithm~1. This approach solves two subproblems in an alternating fashion, as elaborated in Section \ref{P_A}. Here, $v>0$ represents the gap of objective function value between two adjacent iterations in outer loop and the algorithm will converge when $v$ is below a predefined threshold $\varepsilon$. In addition, $\widehat{v}>0$ denotes the penalty violation, $\omega>1$ is the scaling factor of the penalty coefficient. For the proposed algorithm, the main computational complexity comes from solving the standard SDP subproblems in the second subproblem. %, i.e., obtaining the joint passive beamforming solution of STAR-RIS.
For solving  problem \eqref{eq_passive_opt_trans1}, the main  complexity is dominated by $\mathcal{O}(2M^{3.5})$, %Based on the above analysis, the computational complexity of the proposed alternating Algorithm 1 for solving the original covert communication problem  \eqref{eq_ori_opt} is calculated as $\mathcal{O}\left(I_1I_22M^{3.5}\right)$, where $I_1$ denotes the total iteration number in outer loop, $I_2$ represents the total iteration number in inner loop,
which indicates that the complexity is mainly determined by the number of elements at STAR-RIS ($M$).
\vspace{-2mm}
\section{Simulation Results}
In this section, the simulation results are presented to evaluate the covert performance of the proposed STAR-RIS assisted NOMA system. In particular, %in terms of the large-scale path loss model,
we set $\rho_0=-20$ dB, $\alpha=2$, and the distances $d_\mathrm{AR}=100m$, $d_\mathrm{rb}=20m$, $d_\mathrm{rc}=15m$ and $d_\mathrm{rw}=25m$. % are assumed as $100m$, $20m$, $15m$ and $25m$, respectively.
Further, we define $\rho=3$ dB, the noise power $\sigma_{\mathrm{b}}^2=-80$ dBm, $\sigma_{\mathrm{c}}^2=-80$ dBm and $\hat{\sigma}_{\mathrm{w}}^2=-80$ dBm. In the proposed algorithm, the tolerance parameters $\varepsilon$ and $\widehat{\varepsilon}$ are set as  $10^{-4}$ and $10^{-5}$, respectively. In order to highlight the advantages of the proposed scheme, we consider a benchmark scheme with two conventional RIS each having $\frac{M}{2}$ elements to replace the STAR-RIS, where one RIS is the reflection-only RIS and the other is transmission-only RIS. We call this baseline scheme as ``RIS-aided" scheme.
%\vspace{-2mm}
\begin{figure}[ht]
	\centering
	\includegraphics[scale=0.28]{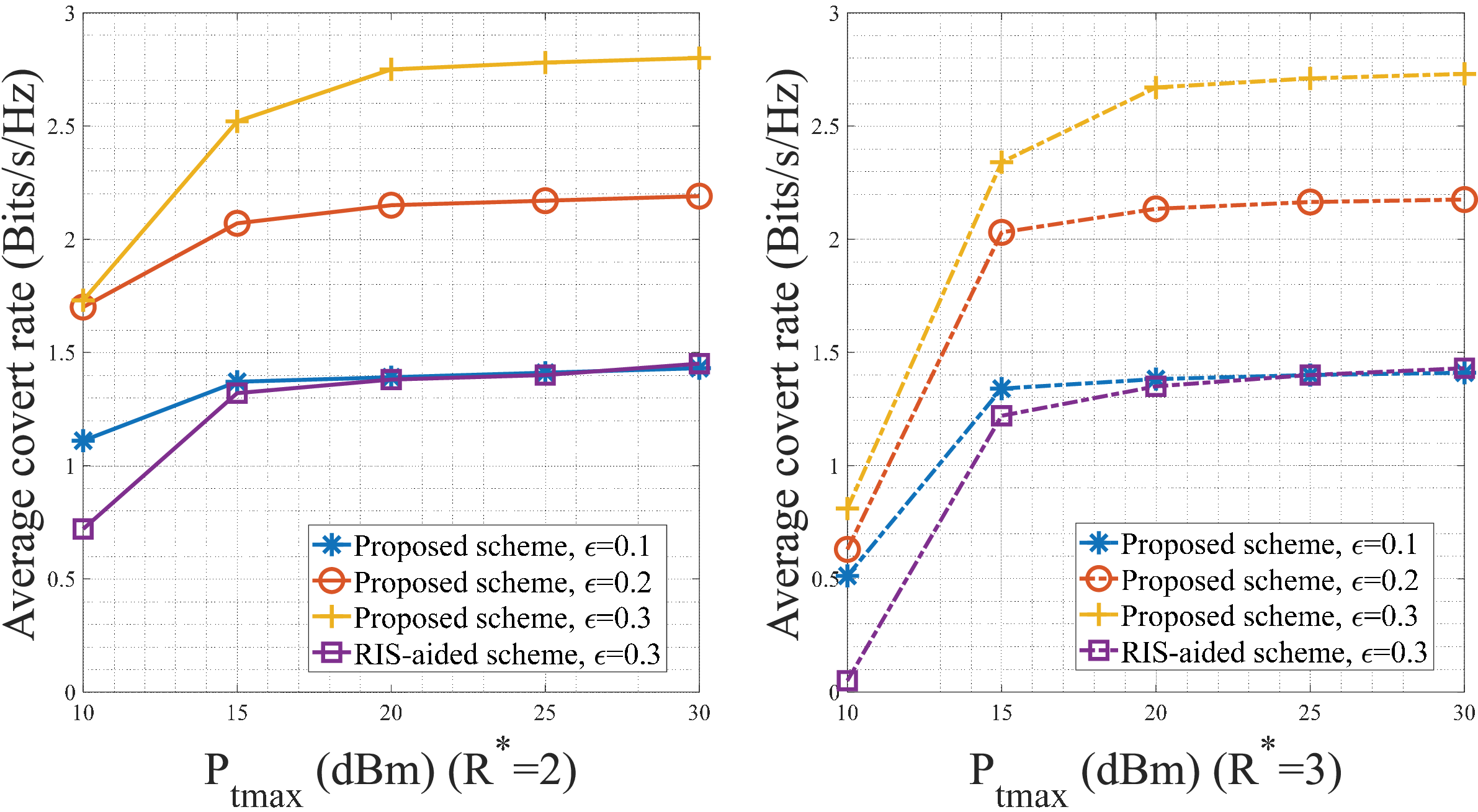}
\vspace{-2mm}
	\caption{Average covert rate versus the maximum transmit power $P_\mathrm{tmax}$.} % with $M=40$, and different covert requirements $\epsilon$ and QoS constraints $R^*$.}
\label{fig:pvsrate}
\end{figure}

Fig. \ref{fig:pvsrate} shows the influence of the maximum transmit power $P_\mathrm{tmax}$ on the average covert rate, considering $M=40$ with different covert requirements $\epsilon$ and QoS constraints $R^*$. Specifically, the covert rate gradually increase with the growth of $P_\mathrm{tmax}$. %, while the increased rate experience an opposite trend for all cases.
In addition, we can find that the higher $\epsilon$ contributes to breaking the performance bottleneck imposed by channel characteristics and the number of elements at STAR-RIS. Compared with the RIS-aided baseline scheme, the proposed scheme possesses a strong superiority in enhancing the covert performance of the system even if tighter covert requirement is adopted. Further, lower covert rates are achieved in tighter QoS constraint (i.e., $R^*=3$) and the degraded performance is the most obvious at $P_\mathrm{tmax}=10$ dBm.

\begin{figure}[ht]
	\centering
	\includegraphics[scale=0.28]{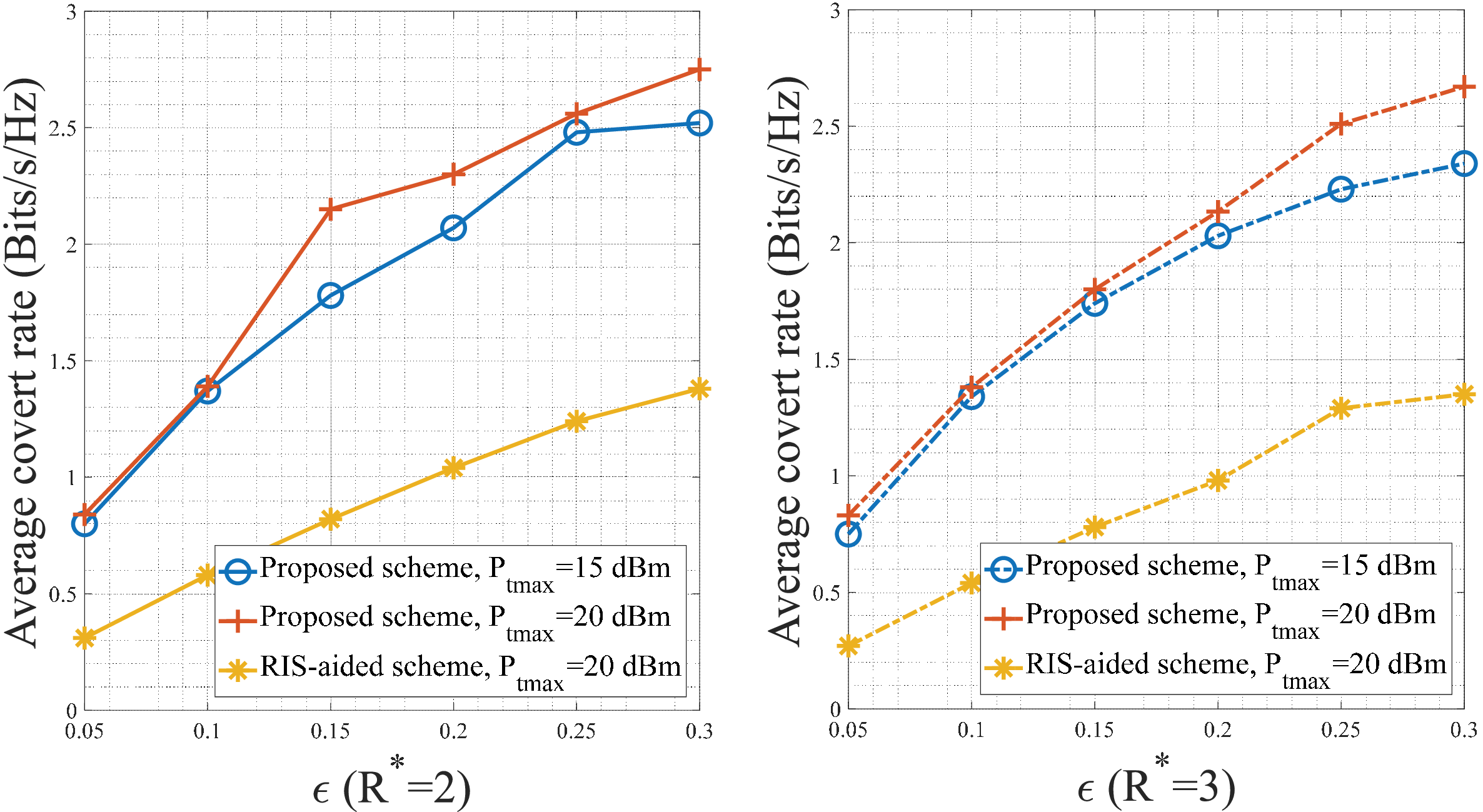}
\vspace{-2mm}
	\caption{Average covert rate versus covert requirement $\epsilon$.} % with $M=40$, and different maximum transmit power $P_\mathrm{tmax}$ and QoS constraints $R^*$.}
\label{fig:epsilonvsrate}
\end{figure}

Next, we explore the performance of average covert rates versus covert requirement $\epsilon$ with different $P_\mathrm{tmax}$ and QoS requirements $R^*$, as presented in Fig. \ref{fig:epsilonvsrate}. %According to the given simulation results,
We can observe that the covert rates have an upward trend with the increase of $\epsilon$ for all cases. To achieve an obvious comparison, $P_\mathrm{tmax}=20$ dBm is selected to implement the RIS-aided baseline scheme, while the acquired performance gain is still far below the proposed scheme even if the proposed scheme is operated at a lower power budget (i.e., $P_\mathrm{tmax}=15$ dBm). This is due to the STAR-RIS offers greater flexibility in reconfiguration as compared to conventional RIS, i.e., it can adjust the element phases and amplitudes for both reflection and transmission.
\vspace{-2mm}
\begin{figure}[ht]
	\centering
	\includegraphics[scale=0.28]{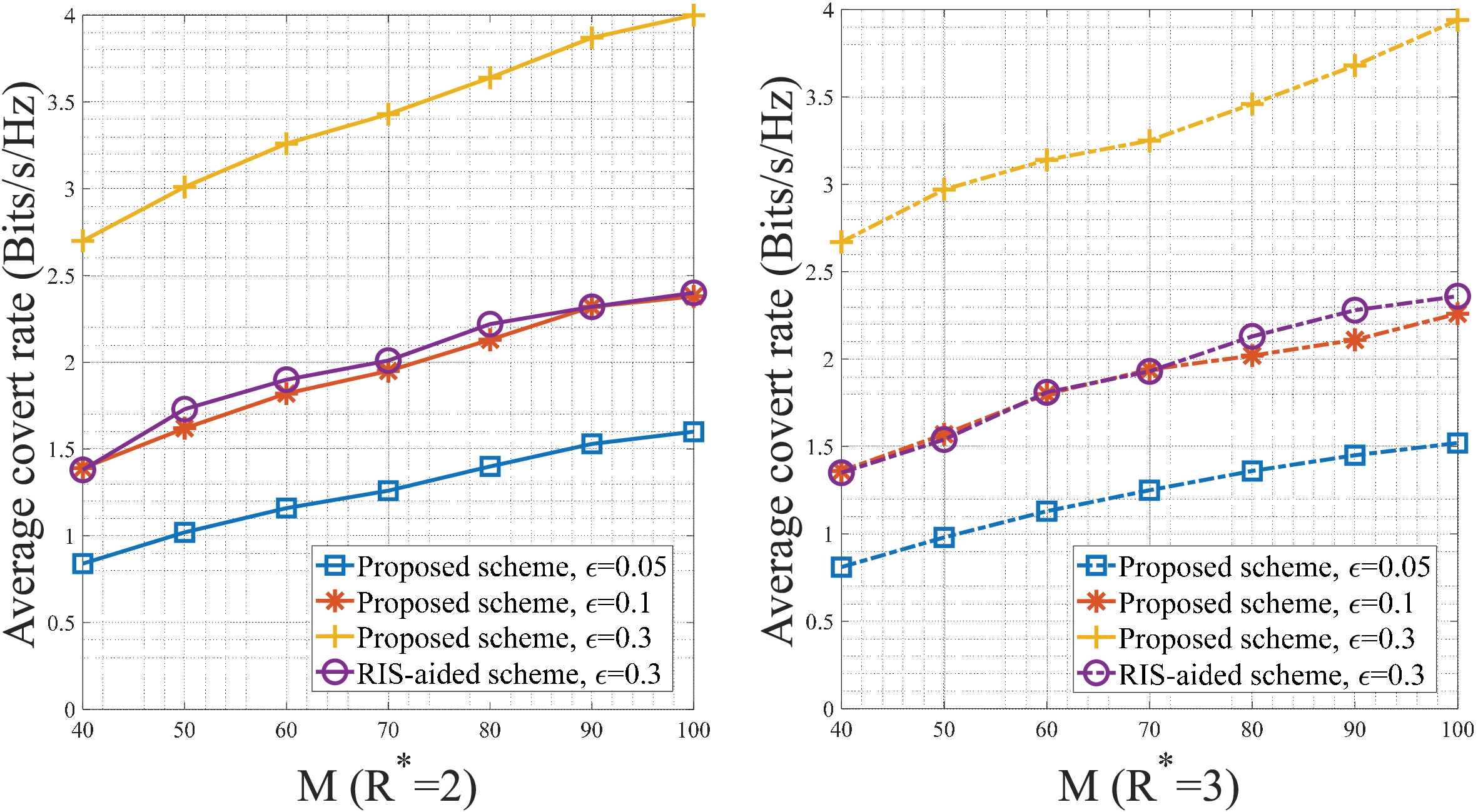}
\vspace{-2mm}
	\caption{Average covert rate versus the element number of STAR-RIS.} %$P_\mathrm{tmax}=20$ dBm, and different covert requirements $\epsilon$ and QoS constraints $R^*$.}
\label{fig:Mvsrate}
\end{figure}

\vspace{-2mm}
In Fig. \ref{fig:Mvsrate}, the variation curves of average covert rate w.r.t. the element number of STAR-RIS ($M$) are shown, under different $\epsilon$ and QoS constraints. It is observed that the average covert rates of all the schemes grow with $M$, since the increased elements can provide higher degrees of freedom for re-configuration of the propagation environment. The RIS-aided baseline scheme is operated with $\epsilon=0.3$. Under the same covert requirement constraint, the proposed STAR-RIS-aided scheme outperforms the benchmark schemes, and the advantage becomes more significant as $M$ increases. %This indicates that the STAR-RIS-aided scheme has a great potential for covert communication.

\vspace{-2mm}
\section{Conclusion}
\vspace{-0.5mm}
In this work, the STAR-RIS-assisted covert communication in NOMA system is investigated. %Specifically,
We first derive the closed-form expression of the minimum DEP which is utilized to characterize the covert performance of the system. Then, an optimization problem maximizing the covert rate of the system under the covertness  and QoS constraints is established by jointly optimizing the transmit power allocation and passive beamformer. Due to the strong coupling among optimization variables, an iterative algorithm  is proposed to effectively solve this optimization problem. Simulation results demonstrate that the STAR-RIS-assisted covert communication scheme highly outperforms the conventional RIS-aided scheme.

\vspace{-2mm}
\appendices
\section{Proof of Theorem 1}\label{Apd_A}
\vspace{-1mm}
According to the derived expression of DEP in \eqref{eq_DEP_expression}, we can see that $P_\mathrm{e}$ is a segment function of detection threshold $\tau_\mathrm{dt}$. Next, let us derive the optimal detection threshold $\tau_\mathrm{dt}^*$.
\subsubsection{$\tau_\mathrm{dt}\geq\phi_2+\frac{\sigma_{\mathrm{w}}^2}{\rho}$} It is easy to derive $P_\mathrm{e}=1-\frac{1}{2\ln\rho}\ln\left(\frac{\tau_\mathrm{dt}-\phi_1}{\tau_\mathrm{dt}-\phi_2}\right)$. First, we calculate the first-order partial derivative of $P_\mathrm{e}$ w.r.t. $\tau_\mathrm{dt}$, which is expressed as
\vspace{-2mm}
\begin{align}	\frac{\partial{P_\mathrm{e}}}{\partial{\tau_\mathrm{dt}}}=\frac{1}{2\ln\rho}\left(\frac{1}{\tau_\mathrm{dt}-\phi_2}-\frac{1}{\tau_\mathrm{dt}-\phi_1}\right).
\end{align}
We can find that $	\frac{\partial{P_\mathrm{e}}}{\partial{\tau_\mathrm{dt}}}>0$ always holds. Hence, the optimal detection threshold in this range is $\tau_\mathrm{dt}^*=\phi_2+\frac{\sigma_{\mathrm{w}}^2}{\rho}$.
\subsubsection{$\tau_\mathrm{dt}<\phi_2+\frac{\sigma_{\mathrm{w}}^2}{\rho}$} The $P_\mathrm{e}$ can be derived as $P_\mathrm{e}=1-\frac{1}{2\ln\rho}\ln\left(\frac{\rho\left(\tau_\mathrm{dt}-\phi_1\right)}{\hat{\sigma}_\mathrm{w}^2}\right)$, and its first-order partial derivative %of $P_\mathrm{e}$
w.r.t. $\tau_\mathrm{dt}$ is given as $\frac{\partial{P_\mathrm{e}}}{\partial{\tau_\mathrm{dt}}}=-\frac{1}{2\ln\rho}\frac{1}{\tau_\mathrm{dt}-\phi_1}$ where $\frac{\partial{P_\mathrm{e}}}{\partial{\tau_\mathrm{dt}}}<0$ holds.

Base on above analysis and $\tau_\mathrm{dt}\in\big[ \frac{\hat{\sigma}_{\mathrm{w}}^2}{\rho}+\phi_1, \rho\hat{\sigma}_{\mathrm{w}}^2+\phi_1\big]$, the optimal detection threshold $\tau_\mathrm{dt}^*$ can be derived as
\vspace{-2mm}
\begin{align}
	\tau_\mathrm{dt}^*=\min\Big\{\phi_2+\frac{\sigma_{\mathrm{w}}^2}{\rho}, \phi_1+\rho\hat{\sigma}_{\mathrm{w}}^2\Big\}.
\end{align}

\ifCLASSOPTIONcaptionsoff %\emph{a}
  \newpage
\fi

\bibliographystyle{IEEEtran}
\bibliography{NOMA-CC}

\end{document}